\documentclass[prd,twocolumn,preprintnumbers]{revtex4-1}

\usepackage{amsmath}
\usepackage{amssymb}
\usepackage{epsfig}
\usepackage{graphicx}
\usepackage{color}
\usepackage[normalem]{ulem}
 \usepackage{url}
\usepackage[breaklinks, plainpages=false, colorlinks=true, anchorcolor=cyan, linkcolor=red, citecolor=cyan, urlcolor=magenta, bookmarks=false]{hyperref}

\usepackage[caption=false]{subfig}

\begin{document}
\renewcommand{\thefigure}{\arabic{figure}}
\setcounter{figure}{0}

 \def\I{{\rm i}}
 \def\E{{\rm e}}
 \def\D{{\rm d}}

\bibliographystyle{apsrev}

\title{Time-frequency analysis for LISA: Fast Waveform Templates}

\author{Neil J. Cornish}
\affiliation{eXtreme Gravity Institute, Department of Physics, Montana State University, Bozeman, Montana 59717, USA}

\begin{abstract}
Time-frequency methods are well suited to the analysis of data from the Laser
Interferometer Space Antenna (LISA), including nonstationary noise and data gaps. Another motivation is that
the sparse time-frequency support of binary signals allows for fast waveform generation, which is the focus of this work.
Amongst the many possible time-frequency representations, the orthogonal Wilson-Daubechies-Meyer transformation
has a frequency translation symmetry that allows for a banded
heterodyne transformation of chirping signals.  When combined with the sparse evaluation of the LISA
time-delay-interferometry (TDI) response using ``TDI on the fly,'' the time-frequency representation has a
small memory footprint and highly parallel operations.  Full-observation,
full-cadence waveforms are avoided: templates are assembled from sparse
carrier tracks and local packet calculations, with short direct endpoint
segments where the slow-track approximation fails.  A family of fast
template-generation algorithms is described for three important source
classes: galactic binaries, massive black-hole binaries, and extreme
mass-ratio inspirals (EMRIs).  The same construction applies to the slowly
evolving portions of all three signals; the principal differences arise at
frequency turnovers, merger, and plunge.  The calculation may be organized
carrier by carrier, or the polarization waveforms may be transformed first and
the detector response applied to their combined time-frequency representation.
The latter construction is referred to as ``TDI Tapestry.''
\end{abstract}

\maketitle

\section{Introduction}

LISA signals are long lived, richly modulated, and often composed of many
harmonics.  A single source may remain in band for months or years, during
which time the orbital motion of the constellation modulates its amplitude and
phase.  The data are also nonstationary.  In particular, the unresolved
Galactic foreground varies over the year as the LISA antenna pattern sweeps
across the Galaxy \cite{Digman:2022jmp}.  Gaps and other local disturbances
are likewise more naturally handled in a time-frequency representation \cite{Pearson:2025wfd}.

A time-frequency basis is a natural choice for LISA analyses~\cite{Cornish:2020odn}.  
It is simple to incorporate a time-varying noise spectrum, gaps remain
localized, and a chirping signal occupies a sparse collection of pixels rather
than an entire time-frequency plane.  The same sparsity can be used during
template generation and likelihood evaluation: it is unnecessary to compute the likelihood in regions where the
waveform has negligible support.  This paper concentrates on constructing
those sparse templates. Time-frequency methods have previously been used for fast LISA
binary analyses and for non-stationary instrumental and confusion-noise models
\cite{Digman:2022igm,Tenorio:2025gci,Du:2025fes}.

The detector response presents a second computational challenge.  LISA does
not directly measure either gravitational-wave polarization.  Rather, they are encoded in the phase
measurements on the laser links, which are combined with time-dependent delays to form the
time-delay-interferometry (TDI) observables~\cite{Tinto:1999yr,Tinto:2002de,Shaddock:2003dj,Vallisneri:2005tdi}.  A direct template calculation
would evaluate the waveform at every delayed time entering every TDI sample.
At a cadence of order one second, a year contains tens of millions of
samples per channel, each requiring many delayed source evaluations.
TDI-on-the-fly instead samples the slowly varying response on a sparse
solar-system-barycenter (SSB) output grid~\cite{Cornish:2025tdi}.  Here the
rapid carrier phase is removed before the remaining complex response
envelope is interpolated in Cartesian form.  This reduces the
response cost by many orders of magnitude, while retaining very high accuracy, even for
unequal and time-varying arms. Combining TDI-on-the-fly with the sparsity of the time-frequency
representation confines source evaluations and WDM calculations to
source-dependent sparse grids, while also significantly reducing the memory
footprint. Several improvements on the original TDI-on-the-fly algorithm are introduced, including a new, general, grid
placement scheme, and a Cartesian treatment of the slowly varying TDI
modulation that avoids inverse tangents and phase unwrapping.

The overall principle is to avoid calculating a full-observation, rapidly oscillating waveform.
Rather, the source model supplies the slowly sampled
amplitude, phase, frequency, and frequency derivative along each carrier track, and from those the
sparse time-frequency representation is constructed directly~\cite{Cornish:2020odn}.
Two main methods for doing this are described here: a time-domain partition
of unity with heterodyned FFTs, and a lookup table based on chirplet--packet
overlaps with the Wilson-Daubechies-Meyer (WDM) filters that map the local track
geometry directly to the WDM coefficients. In Wheeler's
aphoristic style, these may be called ``templates without waveforms'': an
analysis-ready template is produced without the full-cadence waveform over
the entire observation. A numerical stationary-phase approximation (SPA) is
also available where its accuracy is sufficient. The relative cost of these
methods depends on the source and computing architecture.

The per-harmonic approaches are efficient when the power is concentrated in
a few carriers, but their repeated response and packet calculations add up
for models with hundreds of harmonics. For those signals, an alternative
approach reverses the order of operations: the complete summed waveform is
transformed to the WDM basis before the fast TDI response is applied. The
final step generalizes per-harmonic TDI-on-the-fly, which exploits the slow
variation of the response in time. This extension, called ``TDI Tapestry,''
also exploits its slow variation in frequency. It can be faster than the
per-harmonic approach for waveforms with many harmonics, such as binaries
with high orbital eccentricity.

At first sight, galactic binaries, massive black-hole mergers, and
EMRIs appear to require rather different algorithms.  A nearly monochromatic, circular galactic
binary traces one gently curving line; a quasi-circular massive black hole waveform contains a
handful of tracks that turn sharply upward in frequency at merger; and an EMRI can contain
tens or hundreds of precessing harmonics, some of which turn over before the
plunge~\cite{Estelles:2020thm,Chua:2020stf,Katz:2021yft}.  The common structure
is nevertheless more important than these differences.  Each waveform can be
written as a sum of carriers with slowly varying amplitudes and instantaneous
frequencies.  The LISA response is slowly varying across the temporal support of the
time-frequency packet, and the Meyer windows used here are compact in
frequency.  These properties allow the same sparse TDI and WDM machinery to be
used for all three source classes.

The carrier representation, adaptive TDI response, sparse WDM support, and
alternative realizations of ``templates without waveforms'' are described
first. TDI Tapestry is then introduced, wherein the detector response is
applied to the combined WDM-transformed polarizations. The methods are
illustrated throughout using a phenomenological waveform
model for galactic binaries~\cite{Cornish:2005gb,Cornish:2007if}, the
IMRPhenomTHM and its precessing IMRPhenomTPHM extension for massive-black-hole
binaries~\cite{Estelles:2020thm,Estelles:2021tphm}, and the Fast EMRI Waveforms (FEW) model for
extreme-mass-ratio inspirals~\cite{Chua:2020stf,Katz:2021yft}.

\section{Fast LISA response and WDM}

\subsection{A common carrier representation}

The natural starting point is the spin-weight $-2$ spherical-harmonic
decomposition of the complex strain,
\begin{align}
 H(t;\iota,\varphi,\psi)\equiv{}&h_+(t)-\I h_\times(t)\nonumber\\
 ={}&e^{2\I\psi}\sum_{\ell,m}h_{\ell m}(t)
 {}_{-2}Y_{\ell m}(\iota,\varphi).
 \label{eq:spin-weighted-strain}
\end{align}
Here $\iota$ is the inclination of the orbital angular momentum to the line of
sight, $\varphi$ fixes the source-frame azimuth, and $\psi$ is the
polarization angle. Writing an
individual mode as
\begin{equation}
 h_{\ell m}(t)=A_{\ell m}(t)e^{-\I\Phi_{\ell m}(t)},
\end{equation}
separates its slowly varying amplitude from its rapidly evolving phase.

For a nonprecessing binary, reflection symmetry gives
\begin{equation}
 h_{\ell,-m}(t)=(-1)^\ell h_{\ell m}^*(t).
 \label{eq:nonprecessing-mode-symmetry}
\end{equation}
The positive- and negative-$m$ members can therefore be folded into one real
carrier.  Define
\begin{equation}
 {\cal Y}_{\ell m}=e^{2\I\psi}
 {}_{-2}Y_{\ell m}(\iota,\varphi),\qquad q_\ell=(-1)^\ell.
\end{equation}
For $m>0$, the contribution of the pair is
\begin{align}
 H_{\ell|m|}={}&A_{\ell m}
 \left({\cal Y}_{\ell m}e^{-\I\Phi_{\ell m}}
 +q_\ell{\cal Y}_{\ell,-m}e^{\I\Phi_{\ell m}}\right) \nonumber\\
 ={}&A_{\ell m}\left[C_{\ell m}\cos\Phi_{\ell m}
 -\I D_{\ell m}\sin\Phi_{\ell m}\right],
 \label{eq:folded-mode}
\end{align}
where $C_{\ell m}={\cal Y}_{\ell m}+q_\ell{\cal Y}_{\ell,-m}$ and
$D_{\ell m}={\cal Y}_{\ell m}-q_\ell{\cal Y}_{\ell,-m}$.  Consequently,
\begin{align}
 h_+^{\ell|m|}={}&A_{\ell m}
 \left(\Re C_{\ell m}\cos\Phi_{\ell m}
 +\Im D_{\ell m}\sin\Phi_{\ell m}\right),\nonumber\\
 h_\times^{\ell|m|}={}&A_{\ell m}
 \left(-\Im C_{\ell m}\cos\Phi_{\ell m}
 +\Re D_{\ell m}\sin\Phi_{\ell m}\right).
 \label{eq:folded-polarizations}
\end{align}
Thus both polarizations of a $\pm m$ pair share one amplitude, phase, and
time-frequency track; only their cosine and sine projection coefficients
differ.

As an explicit example, consider the dominant $2|2|$ pair. Defining
$c=\cos\iota$, ${\cal N}=\sqrt{5/(64\pi)}$, and
$\phi_{22}=\Phi_{22}-2\varphi$, the two harmonics are
${}_{-2}Y_{22}={\cal N}(1+c)^2e^{2\I\varphi}$ and
${}_{-2}Y_{2,-2}={\cal N}(1-c)^2e^{-2\I\varphi}$.  Equations
\eqref{eq:folded-mode} and \eqref{eq:folded-polarizations} then give
\begin{align}
 h_+^{2|2|}={}&2{\cal N}A_{22}
 (1+c^2)\cos(2\psi)\cos\phi_{22}\nonumber\\
 &+4{\cal N}A_{22}c\sin(2\psi)\sin\phi_{22},\nonumber\\
 h_\times^{2|2|}={}&-2{\cal N}A_{22}
 (1+c^2)\sin(2\psi)\cos\phi_{22}\nonumber\\
 &+4{\cal N}A_{22}c\cos(2\psi)\sin\phi_{22}.
 \label{eq:folded-22}
\end{align}
The implementation uses $\varphi=\pi/2$ to match the $\phi_{\rm ref}=0$ convention of the LIGO analysis library.
This is just a fixed phase-reference choice that can be absorbed into $\Phi_{22}$.

For the response and WDM calculation it is useful to abstract this construction.
Let $P\in\{+,\times\}$ label the two polarizations, and let $\alpha$ index the carriers. The waveforms can then
be written as
\begin{equation}
 h_P(t)=\Re\sum_{\alpha=1}^{N_c}
 {\cal A}_{P\alpha}(t)e^{\I\Phi_\alpha(t)},
 \qquad
 f_\alpha(t)=\frac{1}{2\pi}\frac{d\Phi_\alpha}{dt}.
 \label{eq:carrier-sum}
\end{equation}
The complex amplitudes ${\cal A}_{P\alpha}$ contain the slowly varying
intrinsic mode amplitude and the cosine and sine projection coefficients in
Eq.~\eqref{eq:folded-polarizations}.  The rapid evolution is isolated in the
carrier phase $\Phi_\alpha$.  Precession changes the slowly varying complex
coefficients and broadens each carrier into nearby sidebands, without
requiring a different TDI or WDM interface.

A circular galactic binary has one principal carrier.  The nonprecessing IMRPhenomTHM
model used here combines the positive- and negative-$m$ members into five real
carriers, $(2,|2|)$, $(2,|1|)$, $(3,|3|)$, $(4,|4|)$, and $(5,|5|)$.  FEW
waveforms use combinations of the azimuthal, polar, and radial phases,
\begin{equation}
 \Phi_\alpha(t)=m_\alpha\Phi_\phi(t)
 +k_\alpha\Phi_\theta(t)+n_\alpha\Phi_r(t),
\end{equation}
and generally require many more carriers.  The common representation in
Eq.~\eqref{eq:carrier-sum} simplifies the interface between the waveform model and the TDI response and WDM algorithms.

\subsection{Precessing carrier families}

IMRPhenomTPHM starts from the same five THM mode pairs in a frame that
co-precesses with the binary and rotates them into an inertial frame
\cite{Estelles:2021tphm}.  With the rotation convention used here, the
inertial-frame modes are
\begin{equation}
 h^{\rm I}_{\ell m}(t)=\sum_{m'=-\ell}^{\ell}
 e^{-\I m\alpha(t)}d^\ell_{m'm}\!\left(\beta(t)\right)
 e^{-\I m'\gamma(t)}h^{\rm cop}_{\ell m'}(t),
 \label{eq:tphm-twist}
\end{equation}
where $d^\ell_{m'm}$ is a Wigner small-$d$ matrix and
$\alpha,\beta,\gamma$ are the time-dependent Euler angles.  A single
numerically integrated spin-orbit trajectory supplies these angles for all
the modes; the minimal-rotation condition sets
$\dot\gamma=-\dot\alpha\cos\beta$.  The reflection symmetry in
Eq.~\eqref{eq:nonprecessing-mode-symmetry} is used in the co-precessing
frame, before the rotation.  It does not in general fold a fixed
inertial-frame $\pm m$ pair into one frequency track.

Instead, the positive- and negative-$m'$ members of each co-precessing THM
pair are rotated and projected together.  The result is still represented
by Eq.~\eqref{eq:carrier-sum}, with five parent carrier phases and complex,
slowly varying polarization envelopes ${\cal A}_{P\alpha}(t)$ that contain
the Wigner rotations and observer projection.  The real and imaginary parts
of each envelope are interpolated separately; this remains regular through
sideband cancellations, where an extracted amplitude and phase would develop
a cusp and a jump.  An individual sideband has an instantaneous frequency
shift of approximately $(m\dot\alpha+m'\dot\gamma)/(2\pi)$ from its parent
track.  These shifts widen the occupied band, but need not be transformed as
independent carriers.  The early TDI response is evaluated sparsely for each
of the five complete families, followed by the same partitioned,
heterodyned complex FFT construction used for THM.  The merger and
ringdown use one summed, full-bandwidth endpoint transform per TDI channel.
An SPA on the five folded families would not resolve their separate
stationary points, and unfolding every sideband would discard much of the
computational advantage.

\subsection{Sparse TDI response}

For a TDI channel $I$, the gravitational-wave response can be written
schematically as a finite sum of delayed polarization samples,
\begin{align}
 h_I(t)=\sum_j\big[&
 F^+_{Ij}(t)h_+\!\left(t-\tau_{Ij}(t)\right) \nonumber\\
 &+F^\times_{Ij}(t)h_\times\!\left(t-\tau_{Ij}(t)\right)
 \big].
 \label{eq:tdi-response}
\end{align}
The factors $F^P_{Ij}$ project the wave tensor onto the links, while
$\tau_{Ij}$ contains both the barycenter-to-constellation propagation time and
the appropriate chain of arm delays.  Equation~\eqref{eq:tdi-response} is
valid for unequal and time-dependent arms; no equal-arm transfer function is
factored out of the signal model.

The explicit TDI-1 Michelson response used here contains 16 delayed projected
waveform terms in each of $X$, $Y$, and $Z$, or 48 terms across the three
channels; Eq.~(3) of Ref.~\cite{Cornish:2025tdi} displays all 16 terms for $X$,
with $Y$ and $Z$ obtained by cyclic permutation.  Pairs of these terms share a
retarded source time, allowing the implementation to combine their projection
factors and use only eight distinct waveform evaluations per channel.  The
standard TDI-2~\cite{Shaddock:2003dj} Michelson variables intended for production LISA analyses
double the underlying link sequence, giving 32 delayed projected terms, which
can similarly be grouped into 16 distinct retarded waveform evaluations, per
channel (96 terms or 48 waveform evaluations across $X$, $Y$, and $Z$)
\cite{Vallisneri:2005tdi}.  The larger stencil for TDI-2 increases the response
bookkeeping and cost, but does not alter the sparse-evaluation strategy
described below.

As it stands, Eq.~(\ref{eq:tdi-response}) implies that the waveform must be
sampled many times for each data sample to produce the full set of TDI
observables.  However, the geometric factors that enter the response evolve
mainly on the annual orbital time scale, so the response need not be evaluated
at the data cadence.  The full TDI stencil can instead be sampled infrequently
and used to extract the slowly varying amplitude and phase modulation imparted
by the detector motion, a technique dubbed ``TDI-on-the-fly''
\cite{Cornish:2025tdi}. Below we introduce a more robust and general scheme for setting up the sparse
sample grid, along with a simpler and more robust method for extracting the slow varying amplitude
and phase modulation that is then interpolated to produce the full TDI response.

The time convention that is used in the construction is worth spelling out, as it mixes the fixed SSB frame with
the time varying detector guiding center frame.  Let $t$ be the SSB
output timestamp attached to the data, ${\bf r}_0(t)$ the LISA guiding-center
position, and $\hat{k}$ the propagation vector.  The guiding-center retarded
source argument is
\begin{equation}
 u(t)=t-\hat{k}\!\cdot {\bf r}_0(t),
 \label{eq:center-time-map}
\end{equation}
while an individual TDI term samples the waveform at times
\begin{equation}
 s_{Ij}(t)=t-\hat{k}\!\cdot {\bf r}_{i_j}(t)-\Delta L_{Ij}(t).
 \label{eq:tdi-source-time}
\end{equation}
The WDM pixels and likelihood are labelled by $t$; merger, plunge, and model
transition times are events in the source argument $s$.  Such an event must
therefore be mapped to its output arrival time before it is compared with a
response array.  The time stamps $t$, $u$, and $s_{Ij}$ can differ by tens to hundreds of
seconds, and the waveform can look very different at these times close to merger or plunge.

An adaptive output grid is constructed and the waveform is evaluated at the
corresponding delayed arguments in Eq.~\eqref{eq:tdi-source-time}.  In the THM
implementation trial points are conveniently marched in $u$ and mapped to the
corresponding output time $t$ by inverting Eq.~\eqref{eq:center-time-map}; the FEW
implementation marches directly in $t$ and evaluates all eight exact delayed
arguments in each requested TDI channel.  These are equivalent bookkeeping
choices provided that the coordinate map is retained.  A conservative
orbit-limited step is used during the slow evolution of the binary.  The grid becomes finer
when the carrier frequency or amplitude develops appreciable curvature, when
a delayed TDI term reaches a rapidly evolving part of the signal, or when a
track crosses a TDI delay null.

The spacing is chosen by marching forward from the beginning of the
observation.  A trial interval $[q_0,q_1]$, of width $\Delta q$, is evaluated
at its endpoints and midpoint.  Here $q=u$ in the THM planner and $q=t$ in the
FEW planner; in either case every frequency is evaluated at the mapped
source argument appropriate to that time.  For each carrier let
$f_0=f(q_0)$ and $f_1=f(q_1)$, and define
\begin{align}
 f_s&=\max\left(|f_0|,|f_m|,|f_1|,f_{\rm floor}\right),\nonumber\\
 \delta f_m&=\left|f_m-\frac{f_0+f_1}{2}\right|,
 \qquad f_m=f\!\left(\frac{q_0+q_1}{2}\right).
 \label{eq:tdi-grid-curvature}
\end{align}
An interval is accepted only when every carrier satisfies
\begin{align}
 \frac{|f_1-f_0|}{f_s}&\leq 0.05,
 &\frac{\delta f_m}{f_s}&\leq 0.02,\nonumber\\
 \frac{\pi}{2}\delta f_m\Delta q&\leq 0.25.&&
 \label{eq:tdi-grid-tests}
\end{align}
The first condition limits the frequency excursion across the interval, the
second measures the relative track curvature, and the third limits the phase
error implied by that curvature. An analogous midpoint check limits the
amplitude curvature on the sparse intrinsic source grid.  The frequency tests
are repeated for the reference-carrier argument and at representative TDI-1
delays $0,L,2L,$ and $4L$, so that a delayed copy reaching a rapidly evolving
part of the waveform forces the grid to refine before the undelayed carrier
does.  The actual response still uses the exact unequal-arm arguments in
Eq.~\eqref{eq:tdi-source-time}.  Every retained carrier is tested;
no monotonicity of $f(t)$ is assumed. The default values on the tolerances in Eq.~\ref{eq:tdi-grid-tests}
were arrived at by testing across a wide range of source parameters, and have been
set to conservative values. 

For expensive waveform models the adaptive grid is implemented as two nested meshes.  A
sparse intrinsic grid first identifies where the source ceases to be smooth on
the orbit cadence.  Before that point, intrinsic amplitude and phase are
interpolated onto the fixed-cadence response knots; afterwards, the complete
delayed-probe tests (\ref{eq:tdi-grid-tests}) control every time step.

During slowly varying portions of the signal the step is capped by the orbital sample
cadence, with a default value $10^4$ s.  Values as large as $10^5$ s can be
used, with a small loss in accuracy. A passing step in the grid refinement may grow by at most a factor $1.2$,
while a failing step is bracketed and refined to the largest passing value.
This continuous bracket refinement is important for avoiding unphysical jumps in the waveform support.
Otherwise, accepting the first member of a discrete sequence of reduced steps can make the template and likelihood jump for an arbitrarily
small parameter change.  Additional points are placed when a track crosses a TDI delay zero, where the signed
response amplitude and its phase vary particularly rapidly.

This forward construction improves on the adaptive prescription used in the
original TDI-on-the-fly study~\cite{Cornish:2025tdi}.  That implementation was
tailored to massive black-hole binaries: the grid was seeded at merger and
stepped both forward and backward in time.  The present algorithm requires only
the local carrier tracks and the observation boundaries.  It therefore applies
without change to slowly evolving galactic binaries, non-monotone EMRI
harmonics, and sources whose merger lies before or after the observation.
The final merger and ringdown of a massive black-hole binary remain a special
case because the several delayed TDI terms can sample very different stages of
the rapidly evolving signal.  A short, full-cadence direct-TDI endpoint region
is used there.  Up to that endpoint treatment, the adaptive per-carrier
construction is common to all source classes.

\begin{figure}[t]
 \centering
 \includegraphics[width=0.98\columnwidth]{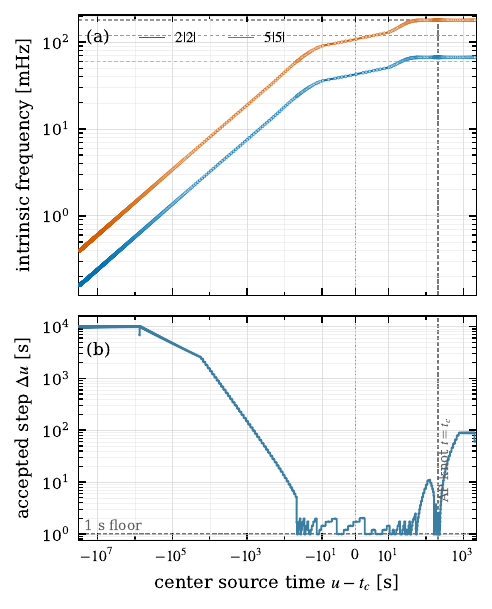}
 \caption{Adaptive response-grid refinement near merger.  Panel (a) shows
 the intrinsic $(2,|2|)$ and $(5,|5|)$ frequencies at the shared response
 knots; connecting lines guide the eye and horizontal dashed lines mark
 $f=n/(2L)$.  Panel (b) shows the accepted center-source interval
 $\Delta u$.  The vertical markers identify the response merger and the
 distinct grid crossing $t=t_c$.}
 \label{fig:thm-adaptive-refinement}
\end{figure}

Figure~\ref{fig:thm-adaptive-refinement} illustrates the resulting refinement
for a higher-mode massive-black-hole binary.  The response knots follow the
slowly varying tracks with a modest cadence until the rapid late-time
evolution and the crossings of the TDI delay nulls require successively
shorter steps.  The same paired response knots are shared by all retained
carriers, so the most demanding mode or delayed probe controls each interval.
The reference source has $m_1=2\times10^5\,M_\odot$,
$m_2=10^5\,M_\odot$, $\chi_1=0.42$, $\chi_2=0.85$, and
$t_c=3\times10^7$~s.  The signed-logarithmic time axis retains the nearly
one-year inspiral while resolving the seconds around the physical
guiding-center merger, $u=t_c$.  For this sky position the response merger
and the SSB-labelled grid location $t=t_c$ are displaced; these are marked by
the dotted and nearby dashed vertical lines.  The orbit-limited $10^4$~s
cadence gives way to source-driven refinement as the track curvature grows,
with additional short intervals at the $f_n=n/(2L)$ delay-null crossings.
Once the frequencies settle to their ringdown values and the amplitudes
decay, the permitted spacing grows again.  A second short refinement about
200~s after $u=t_c$ protects the intrinsic amplitude--phase spline when its
reference-carrier knot crosses $t_c$; it is not a second physical merger.
The separate full-cadence endpoint transform is not shown.

After removing the rapid carrier phase, the slowly varying TDI modulation is represented in Cartesian rather than polar form.
Schematically, if $\Phi_{\rm ref}(u(t))$ is the carrier phase of a harmonic evaluated at the constellation guiding center, and $A(t)$
and $\Phi(t)$ are the amplitude and phase after the TDI response, then we can write
\begin{eqnarray}
h(t) &= &A(t) e^{i \Phi(t)} \nonumber \\
& = & e^{i \Phi_{\rm ref}(u(t))} \left( A(t) \cos\Delta\Phi(t) + i  A(t)  \sin \Delta\Phi(t) \right),
\end{eqnarray}
where $\Delta \Phi(t) = \Phi(t) - \Phi_{\rm ref}(u(t))$. Rather than attempting to extract $A(t)$ and $\Delta \Phi(t)$ directly, which requires inverse
tangents, phase unwrapping, and careful treatment of zero crossings in the amplitude, the slowly varying real and imaginary terms inside the parenthesis are splined directly.

In this Cartesian representation, a TDI cancellation is simply a zero of the
complex response envelope; no signed-amplitude or phase-continuation rule is
needed.  Its real and imaginary components are interpolated with Akima
splines.  Numerical differentiation is a separate issue for the optional
SPA path: local phase and frequency derivatives on the nonuniform response
grid are formed with unequal-step finite differences.  Differentiating an
interpolating spline near a change in knot spacing can introduce false
oscillations, including spurious zeros of $\dot f$.  The partitioned-FFT path
does not require these response-spline derivatives.

The same paired $(u,t)$ response grid can be shared by all carriers.  The constellation
positions, arm vectors, delays, and polarization projections are computed once
at each grid point, after which the different mode phases and amplitudes are
evaluated in vector form.  This sharing is particularly valuable for EMRIs,
where the source may contain dozens of harmonics even after mode pruning.

\subsection{Sparse Meyer--Wilson transform}

An orthonormal Wilson transform built from Meyer windows is used~\cite{Daubechies:1991wv,Necula_2012,Cornish:2020odn,Johnson:2026rrn,Vajpeyi:2026msr}.
The time and frequency spacings obey
\begin{equation}
 \Delta T\,\Delta F=\frac{1}{2}.
\end{equation}
It is useful to first form the complex Meyer packet
\begin{equation}
 Z_{nm}=\int df\,\widetilde h(f)\,
 \widetilde\psi_m^*(f)e^{2\pi\I f t_n},
 \qquad t_n=n\Delta T,
 \label{eq:meyer-packet}
\end{equation}
where $\widetilde\psi_m(f)$ is the compact Fourier-domain Meyer window for
frequency layer layer $f_m= m \Delta F$, and the star denotes complex conjugation.  The real or imaginary
quadrature is then selected according to the Wilson parity~\cite{Daubechies:1991wv}
of $(n,m)$.  The stored transform contains one real coefficient per
time-frequency pixel, while the complex packets retain both quadratures, which is
needed for TDI Tapestry.

The frequency-domain Meyer window is compact.  A carrier therefore contributes
only to layers whose windows intersect $f_\alpha(t)$.  The support planner
advances forward in time and marks all layers touched by the track and the
Meyer half-bandwidth.  This avoids globally inverting $f(t)$ and remains valid
for non-monotone EMRI harmonics.  The support is expanded in time to cover the
time-domain extent of the Meyer window.  Merger and plunge require a separate rule:
the final time strip contains every frequency layer because a short transient
has broad frequency support.

Frequency translation by an even number of WDM layers leaves the packet
calculation invariant, apart from relabeling the layer,
\begin{equation}
\begin{split}
 z_\alpha(t)= A_\alpha(t)e^{\I\Phi_\alpha(t)}
 & \longrightarrow z_\alpha(t)e^{-4\pi\I q\Delta F t}, \\
  m & \longrightarrow m-2q,
 \qquad q\in\mathbb Z. 
 \end{split}
 \label{eq:wdm-heterodyne}
\end{equation}
Thus each carrier or time block can be heterodyned close to zero frequency and
sampled according to its small residual bandwidth rather than its absolute
frequency.  This is the central acceleration used by both the FFT-block and
lookup-table implementations below.

The complex carrier in Eq.~\eqref{eq:wdm-heterodyne} differs from the real
heterodyne used in the original TDI-on-the-fly implementation
\cite{Cornish:2025tdi}, where a slowly varying signal of the form
$A(t)\cos[\Phi(t)-2\pi f_h t]$ was transformed with a real FFT.  A real signal
has conjugate positive- and negative-frequency images, so its translated band
must remain on one side of zero and away from the special WDM zero frequency layer.  The
complex analytic carrier has independent signed Fourier frequencies.  Its
residual track may therefore cross zero, allowing the heterodyne to be placed
near the center of each block and reducing the maximum residual frequency and
required sampling cadence.  The zero bin of this complex FFT represents the
physical frequency $f_h=2q\Delta F$, not the physical WDM zero layer: the block
spectrum is sampled at $f-f_h$ and then assigned directly to the original
interior layer $m$.  The even-layer shift preserves the Meyer window and
Wilson-quadrature parity under this relabeling.

\begin{figure}[t]
 \centering
 \includegraphics[width=0.98\columnwidth]{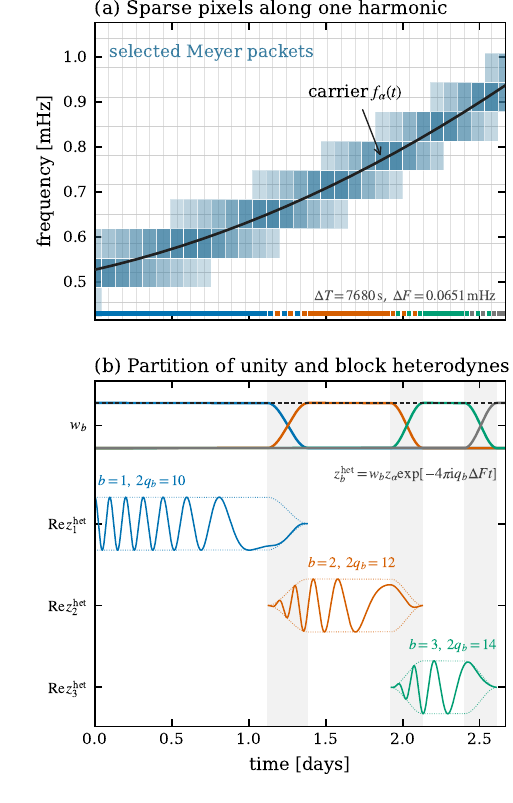}
 \caption{Sparse WDM support and a partitioned block transform for a mildly
 chirping carrier.  Panel (a) shows the selected pixels and the padded block
 supports; dashed alternating colors denote overlaps.  Panel (b) shows the
 complementary windows and the real parts of the three heterodyned blocks.
 The gray member is the endpoint block.}
 \label{fig:wdm-heterodyne}
\end{figure}

Figure~\ref{fig:wdm-heterodyne} shows the two sparse constructions together.
The WDM calculation retains only the packets touched by the carrier and the
compact Meyer frequency window.  The same carrier is split into overlapping
time blocks with complementary tapers that sum to unity.  Each block is heterodyned by an even
multiple of $\Delta F$, so its sampling rate is set by the local frequency
excursion rather than by the absolute carrier frequency.  After the local
FFT, the coefficients are returned to their physical layers by an integer
relabeling.  Solid bars in panel (a) mark the interiors of three early blocks
and the gray endpoint member, while alternating dashed colors mark the shared
overlaps.  The neighboring raised-cosine windows in panel (b) add to unity.
The illustrated blocks use $2q_b=10,12,14$.  Their signed residual frequency
$f_\alpha-2q_b\Delta F$ may begin below zero and rise through it.  The real
part of such a baseband signal can therefore appear to anti-chirp while
$|f_\alpha-2q_b\Delta F|$ decreases, even though the physical track in panel
(a) continues upward.  This visual reversal is a heterodyne convention, not a
change in the source evolution. The apparent anti-chirp of a heterodyned block is analogous to
the ``wagon-wheel'' effect in cinema, where the frame rate of the camera relative to the rotation rate of
a wheel can make it appear that the wheel is rotating in the wrong direction. The difference is that
the cinematic case leads to a loss of information through aliasing, while the frequency reversal in the
complex heterodyne is lossless and invertible.

\subsection{Source-dependent endpoints}

The early evolution has a common treatment, but the endpoint is
source-dependent.  A slowly evolving galactic binary is covered by one
narrow-band heterodyned FFT for the full observation.  Low-eccentricity
generalizations produce closely spaced triplets that can share a heterodyne
and FFT block; an intrinsic negative $\dot f$ from mass transfer requires no
change to the algorithm.

\begin{figure}[t]
 \centering
 \includegraphics[width=0.98\columnwidth]{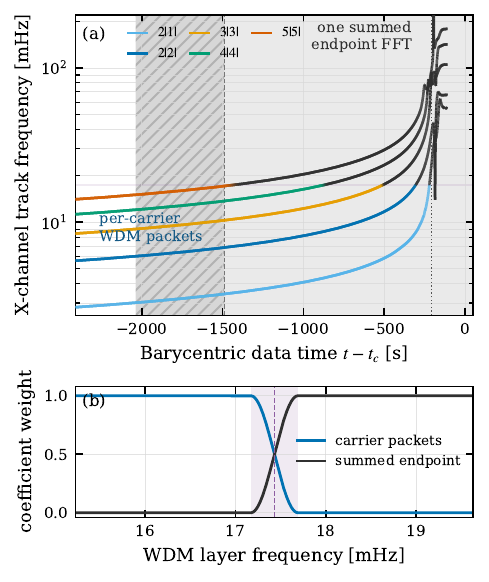}
 \caption{Optional SPA-to-endpoint transition for IMRPhenomTHM in the X channel.
 Panel (a) shows the five
 folded tracks: carrier-owned segments are colored and endpoint-owned
 segments are black.  Hatching and pale shading mark the endpoint taper and
 support; the purple band is the frequency-layer blend.  Panel (b) shows the
 complementary coefficient weights.}
 \label{fig:thm-endpoint-transition}
\end{figure}

For IMRPhenomTHM and IMRPhenomTPHM, the default early path uses partitioned,
heterodyned complex FFT blocks for each folded carrier family.  A numerical
SPA remains available as a faster, less accurate THM option.  With the Fourier convention used in
Eq.~\eqref{eq:meyer-packet}, a stationary time $t_f$ satisfies
\begin{equation}
 f=\frac{1}{2\pi}\dot\Phi(t_f),
\end{equation}
and the leading positive-frequency contribution is
\begin{equation}
 \begin{split}
 \widetilde h(f)\simeq{}&
 \frac{A(t_f)}{2\sqrt{|\dot f(t_f)|}}\\
 &\times\exp\!\left\{\I\left[\Phi(t_f)-2\pi f t_f
 +\frac{\pi}{4}\operatorname{sgn}\dot f(t_f)\right]\right\}.
 \end{split}
 \label{eq:numerical-spa}
\end{equation}
Near merger, where the track steepens and then approaches its ringdown
frequency, the summed higher-mode waveform is evaluated directly through the
TDI delays on a short full-cadence interval.  A single endpoint FFT per channel
then takes over from the early carrier blocks.  The default partitioned path
uses complementary time-domain windows, while the optional THM SPA path uses
a smooth overlap in WDM coefficient space to prevent a discrete layer handoff
from introducing parameter-dependent jumps.  The latter handoff is shown in
Fig.~\ref{fig:thm-endpoint-transition}.  The
frequency center of the established handoff is retained.  Its lower blend edge
is mapped back through every pre-merger carrier track, and the common endpoint
roll-on is shortened so that the time-domain taper is already flat at the
earliest such crossing.  If the existing power-of-two interval cannot begin
early enough, it is enlarged while preserving its late boundary.  Thus the
endpoint is advanced by a track- and WDM-dependent rule rather than by a
source-specific time offset.  The figure uses the barycentric data coordinate
$t-t_c$ that labels the WDM pixels.  The dotted line is the guiding-center
arrival $u(t)=t_c$; its offset from zero is the physical
barycenter-to-detector propagation delay.  In the upper panel, hatching marks
the rising endpoint taper, the pale strip its full time support, and the
purple band the four-layer coefficient blend.  Each track consequently fades
from its carrier color to black.  Carrier ownership terminates at the stored
single-branch SPA stopping time, so a raw post-merger track that turns back
through the blend cannot re-enter the carrier transform.  The display ends
100~s after the response merger because the later phase-derived track has
negligible amplitude, although the endpoint FFT continues through ringdown.
This frequency-layer illustration applies to the optional SPA branch;
the default partitioned FFT joins its carrier and endpoint pieces with
complementary time-domain windows.

\begin{figure}[t]
 \centering
 \includegraphics[width=0.98\columnwidth]{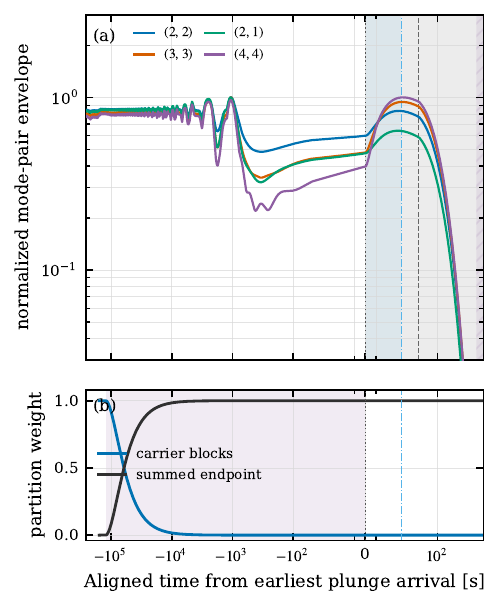}
 \caption{EMRI transition from per-carrier packets to the common plunge
 transform.  Panel (a) shows the four strongest normalized $(\ell,|m|)$
 mode-pair envelopes.  The dotted, dash-dotted, and dashed lines mark the
 earliest delayed arrival, latest arrival, and end of the $C^1$ bridge;
 hatching marks the final fade.  Panel (b) shows the complementary carrier
 and endpoint weights on the shared signed-log time axis.}
 \label{fig:emri-endpoint-transition}
\end{figure}

EMRI harmonics can turn over at widely separated times.  A conventional SPA
is singular at $\dot f=0$ and has two stationary points on the two sides of a
turnover.  In the WDM representation, however, only packets whose time support
straddles the turnover need both branches.  Each such region is covered with a
small, tapered, heterodyned FFT.  The early and late pieces use complementary
windows, so they sum to the original signal rather than being joined at a
single time.  The plunge is handled by one common all-mode endpoint FFT.  The
present FEW waveforms terminate at the end of the inspiral; to avoid the broad
spectral leakage of a hard cutoff, a short $C^1$ quasinormal-mode
completion is attached before the endpoint transform.  The partitioned FFT is the default
EMRI construction.  An SPA path remains available, but in the examples studied
here it has nearly the same cost and lower accuracy.
The reference example is a one-year equatorial Kerr EMRI with
$M=10^6\,M_\odot$, $\mu=50\,M_\odot$, $a=0.5$, and initial eccentricity
$e_0=0.4$.
Figure~\ref{fig:emri-endpoint-transition} shows that the corresponding EMRI
handoff is a partition in TDI data/detector time rather than a blend between
frequency layers.  All early carrier blocks use the complementary falling
window, while one summed endpoint waveform uses the rising window.  The need
for the endpoint transform is not apparent from the nearly horizontal
frequency tracks on this timescale.  It is instead set by the terminal
amplitude evolution: the FEW inspiral amplitudes end at plunge. This abrupt termination
leads to severe and unphysical spectral leakage. To correct this behavior, the
appropriate Kerr quasi-normal ringdown waveform is attached to each mode. The attachment is continuous 
and once differentiable $(C^1)$.
A shared signed-log axis is used in Fig.~\ref{fig:emri-endpoint-transition} to place the day-long
partition taper and the $O(10M)$ ringdown evolution on the same scale.
Source-time offsets in the amplitude panel are translated so that the FEW
endpoint coincides with the earliest TDI output-time arrival.  The dotted line
marks that arrival; the blue dash-dotted line and narrow band mark the latest
arrival and hence the spread among exact TDI paths.  The dashed line marks the
end of the $10M$ $C^1$ ringdown attachment, and the hatched strip is the terminal
numerical fade.  

\subsection{Templates without waveforms}

A sparse WDM template should not be obtained by first generating a dense
waveform and then discarding almost all of its transform coefficients.  Two
constructions avoid that intermediate object.  In the partitioned-FFT
construction, each carrier is divided into overlapping time intervals.
Complementary rising and falling tapers form a partition of unity, so the sum
of the windowed pieces is exactly the original finite-duration signal.  Within
each interval the carrier is heterodyned close to baseband and sampled at a
rate set by its local bandwidth rather than by its absolute frequency.  Only
these short, slowly oscillating pieces are formed.  Their FFTs are then
projected onto the required Meyer packets.  This method is numerically
faithful apart from controlled sampling and taper errors, and it makes
effective use of highly optimized FFT libraries.

The lookup-table construction, introduced in Ref.~\cite{Cornish:2020odn},
takes the idea one step further: even the short baseband waveform need not be
formed.  Near the center $t_n$ of a WDM packet, a carrier is represented
locally as
\begin{align}
 z_\alpha(t_n+\tau) \simeq {}&
 \left({\cal A}_{\alpha n}+\dot{\cal A}_{\alpha n}\tau\right)
 e^{\I\Phi_{\alpha n}} \nonumber\\
 &\times \exp\!\left(2\pi\I f_{\alpha n}\tau
 +\I\pi\dot f_{\alpha n}\tau^2\right).
 \label{eq:local-chirplet}
\end{align}
If $f_c$ and $\dot f_c$ denote the instantaneous frequency and chirp rate at
the pixel, the natural lookup coordinates are
\begin{equation}
 u = \frac{f_c-m\,\Delta F}{\Delta F}, \qquad
 v = \frac{\dot f_c T_{\rm w}}{\Delta F},
\end{equation}
where $m$ is the frequency-layer index and $T_{\rm w}$ is the effective duration of the Meyer window. Schematically, the tabulated moments are
\begin{align}
 K_q(u,v)={}&\int d\tau\,\left(\frac{\tau}{T_{\rm w}}\right)^q
 \psi_m(\tau) \nonumber\\
 &\times\exp\!\left\{2\pi \I u\Delta F\tau
 +\I\pi\frac{v\Delta F}{T_{\rm w}}\tau^2\right\}.
\end{align}
Up to the WDM normalization and Wilson-quadrature conventions, the carrier's
pixel coefficient is assembled directly as
\begin{equation}
 W_{nm}^{(\alpha)} \propto e^{\I\Phi_{\alpha n}}
 \left[\mathcal{A}_{\alpha n}K_0(u,v)
 +T_{\rm w}\dot{\cal A}_{\alpha n}K_1(u,v)+\cdots\right].
 \label{eq:lookup-pixel}
\end{equation}
Thus the calculation consumes only the amplitude, phase, frequency, and
frequency derivative at the packet center.  No time series is stepped through
its cycles and no global Fourier series is built.  The $q=0$ table supplies
the locally constant-amplitude result, while the first moment gives an
inexpensive correction for a linear amplitude variation across the packet.
The table is symmetric in $v$ and therefore treats rising and falling tracks
equally. This is important not only for EMRI mode turnovers, but also for
galactic binaries with negative $\dot f$ caused by mass transfer.

The two algorithms have similar CPU costs in the present galactic-binary and EMRI examples, despite performing very different operations. The block method pays for padded samples, overlapping intervals, and FFT setup, but benefits from contiguous memory access. The lookup method works directly on the active WDM pixels and avoids FFT planning, but pays for multidimensional interpolation, complex phase factors, and the Wilson-quadrature projection at every pixel. Consequently there is no architecture-independent winner. A small number of compact carriers favors the simple FFT construction, while a large or fragmented set of tracks can favor the lookup table. Batched likelihood calculations on GPUs or other vector architectures may particularly favor the regular, independent per-pixel lookup kernel, provided the table remains small enough for fast memory. Conversely, optimized CPU FFT libraries can make fairly large blocks inexpensive. The choice will depend on the available hardware and how the waveforms and likelihoods are used in the analysis. The partitioned FFT is the conservative EMRI default because it currently provides the higher accuracy.

\subsection{Spin Precession}

Spin precession causes the orbital plane to tilt, which changes the instantaneous inclination and polarization angles. In terms of the spin weight -2 spherical harmonics, spin precession
populates the full list of $m$-modes in the detector frame. One might wonder if this unduly complicates the sparse TDI and WDM construction. Thankfully, it does not. The precession side-bands
are very narrow - of order micro Hertz for typical LISA systems, and the time-frequency content of precessing and non-precessing systems is almost identical.

\begin{figure}[t]
 \centering
 \includegraphics[width=0.98\columnwidth]{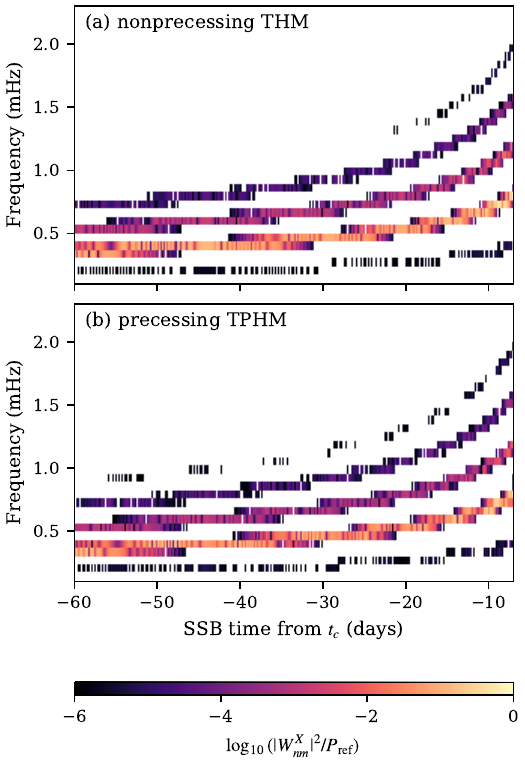}
 \caption{Unwhitened X-channel WDM power for nonprecessing THM (a) and
 precessing TPHM (b), with the same masses, spin magnitudes, viewing
 angle, and extrinsic parameters.  Both panels use the same power scale;
 white pixels are unpopulated or below $10^{-6}P_{\rm ref}$.}
 \label{fig:thm-tphm-wdm}
\end{figure}

\begin{figure}[t]
 \centering
 \includegraphics[width=0.98\columnwidth]{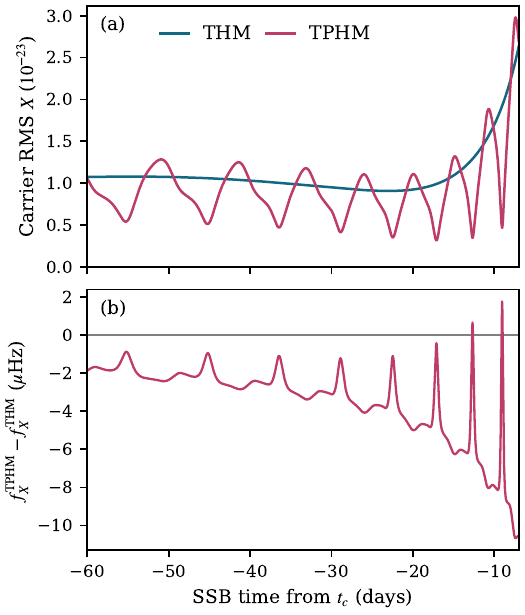}
 \caption{X-channel TDI response for the THM and TPHM sources of
 Fig.~\ref{fig:thm-tphm-wdm}.  The upper panel shows the full-mode
 carrier RMS; the lower panel shows the difference between the summed
 TPHM and THM $2|2|$ X-response frequencies.}
 \label{fig:thm-tphm-response}
\end{figure}

As an example, consider a massive black hole binary with masses $2.4\times10^5+0.6\times10^5\,M_\odot$ viewed at $\iota=\pi/4$.  Both spin magnitudes are $0.95$.  In
the nonprecessing case $\chi_{1z}=0.95$ and $\chi_{2z}=-0.95$, while the precessing reference spins are $\boldsymbol\chi_1=\boldsymbol\chi_2=(0.95,0,0)$.  All other source
parameters are identical.  The one-year observation has $t_c=3.0\times10^7\,{\rm s}$,
$D_L=1\,{\rm Gpc}$, ecliptic colatitude and longitude $(2.31,0.57)$,
polarization angle $0.4$, and reference phase zero.  The precessing spins
are specified at the initial source-time boundary of the observation.
Figure~\ref{fig:thm-tphm-wdm} compares the unwhitened X-channel WDM power
from 60 to 7 days before merger.  Fig.~\ref{fig:thm-tphm-response} shows that precession leads to amplitude modulation and weak
sideband power, which does not significantly enlarge the time-frequency region covered by the signal.
For this source the full-observation sparse X-channel transforms contain almost exactly the same number of active WDM pixels
for THM and TPHM (both around $10^5$).  This counts depends
on the details of the source and WDM setup, but the difference in time-frequency volume caused by precession is generally very small.

The upper panel of Fig.~\ref{fig:thm-tphm-response} displays
$X_{\rm rms}(t)=[\sum_\alpha |X^{\rm an}_\alpha(t)|^2/2]^{1/2}$,
where $X^{\rm an}_\alpha$ is the complex quadrature of one complete
carrier after the TDI delays, summed over all five parent families.  By using this
phase-averaged measure we can see the precession modulation without the distraction of
millions of unresolved cycles.  The lower panel compares the
instantaneous frequencies of the THM and TPHM $2|2|$ X responses.  For
TPHM the full folded $2|2|$ family is summed before its phase derivative
is taken; the small excursions arise from interference between its
precession sidebands.  It is this interference that would break the standard stationary phase
approximation~\cite{Klein:2013qda,Chatziioannou:2017tdw}. The rapidly varying parent phase is removed
before numerical differentiation and its frequency is restored
afterward.  The response-frequency difference remains small compared
with the overall chirp, consistent with the similar time-frequency
footprints in Fig.~\ref{fig:thm-tphm-wdm}.

For the same precessing source, a direct TDI evaluation at $1.875\,{\rm s}$
cadence, followed by a full-length FFT and WDM transform, provides a
reference for the fast construction.  Both paths use a $10^5\,{\rm s}$
start taper.  The normalized, unwhitened WDM coefficient matches, without
time or phase maximization, are $0.9999850$, $0.9999929$, and $0.9999846$
for X, Y, and Z.  The sparse support captures more than $99.99997\%$ of
the direct-reference power in each channel.  The computational speed up on the same single CPU core was a
factor of just under two thousand. Further speed ups could come from a well designed GPU implementation.
The sparsity of the fast TDI+WDM approach leads to a small memory footprint, which can be advantageous for GPU
architectures.


\section{TDI Tapestry}

The carrier-by-carrier construction applies TDI and then the WDM transform to
each term in Eq.~\eqref{eq:carrier-sum}.  This is efficient for one galactic
binary or the five folded IMRPhenomTHM/TPHM carriers, but its cost grows with both
the number of carriers and the number of detector channels.  For a many-mode
EMRI, repeating nearly the same slowly varying detector response for every
track becomes expensive.  TDI Tapestry reverses these operations during the
adiabatic part of the signal: the carriers are first woven into two
polarization time-frequency fields, after which the response is applied once
to the occupied pixels. TDI and the WDM transform do not commute exactly;
Tapestry approximates their composition by a local response in the
time-frequency plane, with corrections for variation across each packet.

To see how this works, approximate each delayed term in
Eq.~\eqref{eq:tdi-response} locally by a monochromatic signal at $(t_n,f_m)$.
The complex TDI symbol for polarization $P$ is
\begin{equation}
 R_I^P(t_n,f_m)=\sum_j F^P_{Ij}(t_n)
 \exp\!\left[-2\pi\I f_m\tau_{Ij}(t_n)\right].
 \label{eq:tdi-symbol}
\end{equation}
The precise bookkeeping of the time origin can move a common phase between the
symbol and the waveform packet, but leaves their product unchanged.  All arm
lengths and projection factors in Eq.~\eqref{eq:tdi-symbol} are evaluated from
the time-dependent constellation; the frequency dependence of every delay is
analytic.

Let $C_P(n,m)$ denote the complex Meyer packet of polarization $P$.  With
${\cal P}_{nm}$ selecting the Wilson-parity quadrature, define
$w_P={\cal P}_{nm}[C_P]$ and its companion
$\widetilde w_P={\cal P}_{nm}[\I C_P]$.  At leading order the detector coefficient is
\begin{equation}
 \begin{split}
 w_I(n,m)={}&\Re R_I^+\,w_+ +\Im R_I^+\,\widetilde w_+\\
 &+\Re R_I^\times\,w_\times
  +\Im R_I^\times\,\widetilde w_\times .
 \end{split}
 \label{eq:tapestry-leading}
\end{equation}
Retaining both quadratures is essential.  A standard real Wilson transform
stores only one of them at each pixel and does not contain enough information
to apply the complex response after the transform.

The response is not perfectly constant across a Meyer window.  Expanding it
in frequency about the layer center gives a controlled correction.  Define
the polarization moments
\begin{equation}
 C_P^{(q)}(n,m)=\int df\,
 \left(\frac{f-f_m}{\Delta F}\right)^q
 \widetilde h_P(f)\widetilde\psi_{nm}^*(f).
 \label{eq:tapestry-moments}
\end{equation}
The response through order $Q$ is then
\begin{equation}
 \begin{split}
 W_I^{(Q)}(n,m)={\cal P}_{nm}\Bigg[
 &\sum_P\sum_{q=0}^{Q}\frac{\Delta F^q}{q!}\\
 &\times\partial_f^qR_I^P(t_n,f_m)C_P^{(q)}(n,m)
 \Bigg],
 \end{split}
 \label{eq:tapestry-expansion}
\end{equation}
where ${\cal P}_{nm}$ selects the appropriate real Wilson quadrature.  The
derivatives of Eq.~\eqref{eq:tdi-symbol} are inexpensive analytic derivatives
of its delay phasors.  In the tests performed here, just keeping the linear term in the expansion reduces
the Tapestry mismatch from order $10^{-3}$ to order $10^{-5}$.  The quadratic
term brings little additional improvement, while increasing the transform and
memory cost.  The first-order frequency correction is therefore retained as
the default.

\begin{figure}[t]
 \centering
 \includegraphics[width=0.49\textwidth]{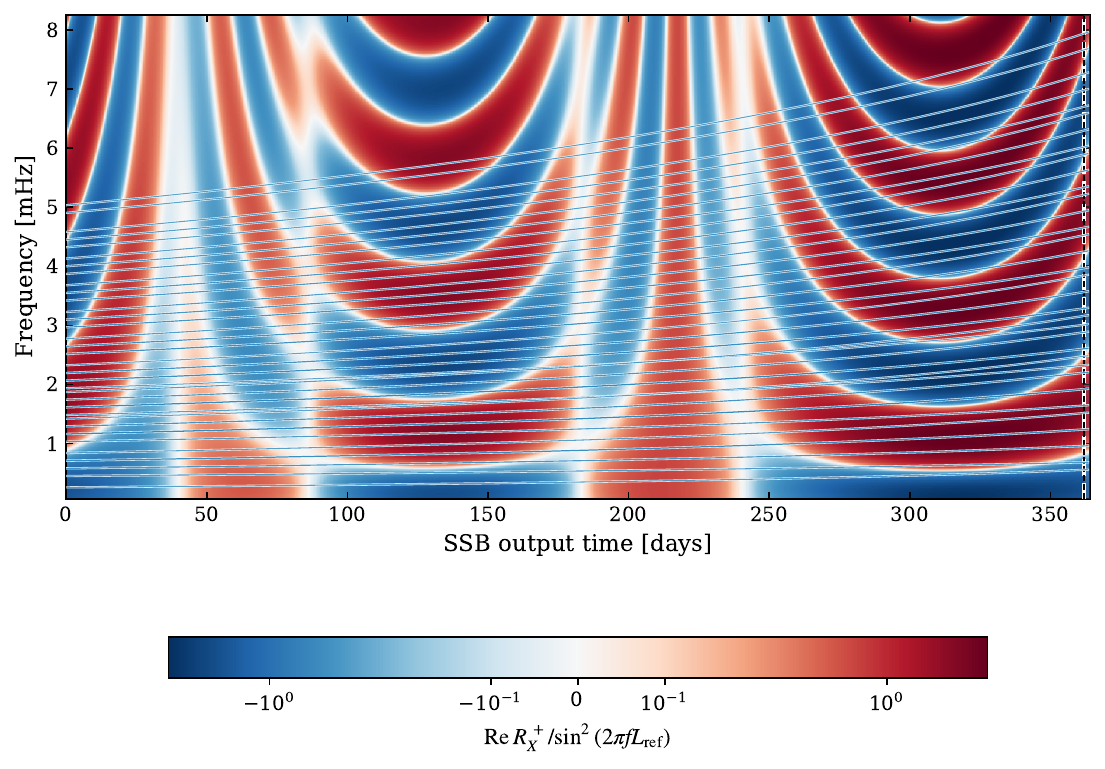}
 \caption{Signed real X-channel plus response, reduced by
 $\sin^2(2\pi fL_{\rm ref})$, for the one-year equatorial Kerr EMRI.
 Blue curves are the 34 retained carriers; the dashed line
 marks the overlap with the direct endpoint construction.}
 \label{fig:tapestry-response}
\end{figure}

Figure~\ref{fig:tapestry-response} shows the real Cartesian component of the
X-channel plus-polarization response for the reference equatorial Kerr EMRI.
The leading equal-arm factor $\sin^2(2\pi fL_{\rm ref})$ is divided out only
for display; the calculation uses the full complex response.  Unlike a
response phase, this signed component remains defined through its zero
crossings.  It varies smoothly, but nontrivially, in both time and frequency
and can be reused by many carriers.  The 34 exact phase carriers selected for
this source sample different parts of the same field; their line weights
indicate relative strength.  The dashed line and pale terminal strip identify
the overlap with the direct endpoint construction, where the summed plunge
and QNM completion are evaluated with the exact delayed TDI response.

Tapestry is used only where the response is adiabatic across a packet.  The
massive-black-hole merger and the EMRI plunge are still formed by summing the
physical time-domain waveform, applying every TDI delay directly, and taking
one short endpoint FFT per channel.  This exact terminal member is joined to
the early Tapestry member by the same partition of unity used in the
carrier-by-carrier construction.  Consequently the approximation in
Eq.~\eqref{eq:tapestry-expansion} is not asked to describe a nearly vertical
track or a broadband endpoint.

The resulting scaling differs from the thread construction.  The expensive
polarization packet transforms are performed twice, rather than once per
carrier and channel, and the sparse response is evaluated once per occupied
pixel and channel.  For the five folded IMRPhenomTHM/TPHM carriers, this is close to
the crossover at which the two methods have comparable cost.  The advantage
grows for EMRIs or eccentric binaries with many harmonics.  Tapestry also
produces a nicely batched calculation, making it a natural candidate for GPU
likelihoods.  The carrier-by-carrier path remains valuable as an independent
reference and can be faster for a small number of narrow tracks.

\section{Future directions}

 The fast TDI and WDM transforms greatly reduce the cost of producing waveform templates for LISA analyses. But the cost is not negligible, and a full LISA global fit~\cite{Cornish:2005qw,Littenberg:2023xpl} will require billions of waveform evaluations. One way to reduce the cost of the analysis is to reuse some of the
 calculations by separating out the ``shape'' and ``projection'' parameters. Sometimes these are referred to as ``intrinsic'' and ``extrinsic''
 parameters, but the time-dependent response of LISA blurs the intrinsic/extrinsic distinction. In terms of the spin-weight $-2$ spherical harmonics, the shape parameters
 determine the form of each harmonic, while the projection parameters change how the harmonics are combined together. 
 
 The sparse TDI and WDM transforms can be applied to each  spin-weighted mode separately rather than
being folded with its negative-$m$ partner, and these transforms can be reused multiple times by adding them together with weightings set by
 the projection parameters. Retaining both packet quadratures for
each mode would allow distance, inclination, polarization, and reference
phase to be varied by recombining stored mode-level responses, rather than
repeating their TDI and WDM transforms.  For a fixed shape and detector
response, the required data--mode and mode--mode inner products would only
involve the union of mode-level active pixels.  This has a close parallel in the
shape--projection split and precomputed inner products of QuickCW
\cite{Becsy:2022quickcw}.  Related intrinsic--extrinsic factorizations are
used by RIFT and cogwheel, including treatments of higher modes and
precession \cite{Lange:2018rift,Islam:2022factorized,Roulet:2024marginalization}.
The matrix-product evaluation of many intrinsic--extrinsic combinations in
dot-PE suggests another way to batch the sparse inner products
\cite{Mushkin:2025dotpe}.

Not every LISA parameter can be moved into a fixed projection coefficient:
sky position changes the projected delays, while merger time shifts the signal
relative to the time-dependent detector response. For precessing sources,
constant rotations setting the source orientation can be treated as projection
parameters, but a time-dependent precession rotation does not generally commute
with the TDI delays or act pointwise on WDM coefficients. For slowly precessing
systems it might be possible to perform the ``twisting up'' procedure after
TDI+WDM if the rotation varies little across the relevant delays and wavelet
support, but this approximation remains to be tested.

 \section*{Acknowledgments}
This work was supported by NASA LISA Preparatory Science Grant 80NSSC19K0320. 

\bibliography{refs}

\end{document}